\documentstyle [12pt] {article}
\begin{document}
\title{DILEPTON PRODUCTION\\ IN ELEMENTARY PROCESSES\\ ON THE NUCLEON
       \thanks{Work supported by BMFT and GSI Darmstadt}}
\author{U. MOSEL\thanks{talk based on work with H.C. D\"onges and M.
        Sch\"afer}\\
        Institut f\"ur Theoretische Physik, Universit\"at Giessen\\
        D--35392 Giessen, Germany}
\date{}
\maketitle
\begin{abstract}
We study elementary processes for production of dileptons from
nucleons,
either through NN bremsstrahlung or through photon-induced reactions.
We
emphasize the dependence of the expected cross-sections on the
electromagnetic formfactor of the nucleon in the time-like regime and
point out that the mentioned reactions can provide important
information
on the validity of vector meson dominance for the nucleon. We also
study
the off-shell dependence of the formfactors.
\end{abstract}

\section{Introduction}

Vector meson dominance is a theoretically well developed concept
for the description of the coupling between hadrons and photons.
While the original idea of Sakurai \cite{Sakurai} relied to a large
extent on
the equality of the quantum numbers of the $\rho$ meson and the
photon and the similarity of the Lagrangians describing both
particles, the present-day picture is that of charged
quark-antiquark loop insertions in the photon propagator which
also can change the interaction vertices.

This picture, appealing
as it is, is experimentally well established only for the pion
where the measured formfactor shows a clear resonance at the
$\rho$ meson mass and can be described by a monopole fit
\cite{Weise}. For
other vector mesons and for the nucleon the picture is much less
clear. For example, for the nucleon the formfactor in the
space-like region exhibits a well-established dipole form, and the
time-like region below a momentum of around 2 GeV/c, i.e. the
vector meson pole region, is experimentally not accessible
\cite{Dubnicka}.
In this paper we argue that the production of dileptons in
elementary processes, e.g. in nucleon-nucleon collisions or in
photon-nucleon collisions can shed some light on the structure of
the electromagnetic formfactor of the nucleon in the time-like
region, and thus on the question whether vector meson dominance
is a useful concept for the nucleon.

This article is based on 3 very recent publications which contain
all the details of the theory and the calculations
\cite{Schaefer1,Doenges,Schaefer2}. The present article therefore
summarizes only the main results without going into any of the
technical
details which can be found in the references just given.

\section{Dilepton Bremsstrahlung}

Dilepton spectra measured in heavy-ion collisions are dominated
by two components, the $\eta$ Dalitz decay and the $\pi^+\pi^-$
annihilation \cite{Wolf,Ko}. Below these two dominant components
lies,
 however,
a strong background of dileptons from nucleon-nucleon
bremsstrahlung and $\Delta$ decay that has a rather strong
bombarding energy dependence \cite{Wolf}. In proton reactions on
nuclei these
two components become dominant for higher invariant masses of the
dileptons. It is, therefore, of interest to study these
components in a more refined model than the commonly employed
long-wavelength approximation.

In \cite{Schaefer1} we have therefore studied the dilepton production
in
nucleon-nucleon collisions using an effective $T$ matrix that is
based on a One-Boson-Exchange (OBE) model; the $T$ matrix is
determined by fitting the nucleon-nucleon elastic scattering. The
calculation treats the dilepton emission from nucleon lines
and from $\Delta$ lines coherently, whereas in standard BUU
simulations these two decay channels do not interfere. The
interference turns out to be quite important at the higher
invariant masses and can describe the observed mass- and
bombarding energy dependence of the $pd/pp$ ratio in the
dilepton yield (see Fig. 1).
\begin{figure}
\vspace{9.5cm}
\caption{Ratio of dilepton invariant mass spectra for pp to pd at
various
bombarding energies from 1.03 to 4.90 GeV. The solid dots give the
data
whereas the line gives results our calculation (from
\protect\cite{Schaefer1}).}
\end{figure}

These calculations were done without any formfactor for the
nucleons; under the assumption of universal VMD the formfactors are
indeed expected to drop out from the ratio. The absolute
cross-sections are, however, very sensitive to the formfactor
used. In our calculation we have simply assumed that the electric
formfactor has, first, a VMD shape and, second, no off-shell
dependence. Under these assumptions the calculations, which are
gauge-invariant, give a strong vector meson signal at the $\rho$
meson mass. There are, so far, no published data on dilepton
production in $pp$ collisions, but under the assumption that a
proton-induced reaction on a light target proceeds mainly through
independent $pp$ and $pn$
collisions, the calculations can -- properly weighted with the
correct number and type of nucleon-nucleon collisions -- be compared
to the $p + {}^9Be$ data obtained by the DLS group at Berkeley (see
Fig. 2).
\begin{figure}
\vspace{6cm}
\caption{Dilepton invariant mass spectrum for $p + {}^9Be$ at 2.1
GeV.
The data are from the DLS group, the lines give results of our
calculations:
no formfactor (dotted), VMD formfactor for nucleon and $\Delta$
(dashed),
formfactor from the model of Rho et al \protect\cite{Rho} (solid)
(from
\protect\cite{Schaefer1}.}
\end{figure}
These data obviously show no special effect in the vector meson
mass region.

\section{Off-Shell Dependence of the Formfactor}

If this result holds up for the $p+p$ data already taken at
Berkeley, but not yet publicly available, then there is obviously
a gross discrepancy between experiment and data and the question
for its reason arises. We have, therefore, studied the
electromagnetic formfactor of the nucleon in a model that
combines a loop expansion for the $\pi$, $N$ and $\Delta$ with
vector meson dominance for the pion \cite{Doenges}. In this picture
the photon couples to the nucleon only through the pion cloud which,
in
turn, is dominated by VMD in its coupling to the photon. The
model allows to study also half-off-shell vertices as they are
important in the bremsstrahlungs studies.

The main result of this study, that reproduces the on-shell
electromagnetic formfactor of the nucleon in the space-like
region quite well, is that the time-like electric formfactor in
the particle-particle channel is
dominated by VMD, and this VMD is remarkably stable as a function
of the off-shellness of the nucleon; the off-shell dependence of
the electric formfactor is thus negligible. This is not true,
however, for the magnetic formfactor which is strongly off-shell
dependent. It is also not true for the antiparticle-particle
channel (see Fig. 3) which can play a significant role in
bremsstrahlungs
processes where -- in the so-called post-emission graphs -- the
nucleon is put off-shell by the NN-interaction before it emits
the dilepton; depending on the structure of the $T$ matrix the
antiparticle admixture in the virtual nucleon line between
interaction and emission can be quite sizeable.
\begin{figure}
\vspace{18cm}
\caption{Electric formfactors for the proton for various off-shell
energies, denoted by the invariant mass $W$, of the incoming proton
(from \protect\cite{Doenges}).}
\end{figure}
It thus seems to be necessary to include these off-shell effects
in any analysis of data on
$p + p$ dileptons with the aim to determine the formfactor,
since for reactions
in the GeV bombarding energy range the nucleon can easily go
off-shell by several hundred MeV.

\section{Compton Scattering into the Timelike Region}

The off-shellness of the nucleon in nucleon-nucleon collisions is
governed by the strong interaction $T$ matrix, so that the
dilepton data are due to an interplay of strong and
electromagnetic vertices. This makes it difficult to extract the
information on the electromagnetic vertex in an unique way.

A ``cleaner'' experiment is the photon-induced dilepton
production $\gamma + p \longrightarrow p + e^+e^-$ that -- at
least in the $s$ and $u$ channels -- is free of any strong
interaction vertex. The $t$ channel, which involves a
photon-photon-meson vertex and a meson-nucleon vertex, does
contribute, but only at forward
angles between the incoming and outgoing photon because of the
low mass of the intermediate virtual pion (in this energy range
there are no other physical mesons with an appreciable photon-vector
meson coupling strength). As we will see
below the forward direction is, however, unsuitable for any
dilepton production experiments, because here the (in this
context uninteresting) Bethe-Heitler contribution dominates. This may
be different if one is interested only in total cross sections for
the
photoproduction of vector mesons \cite{Soyeur}; if these are
identified
by hadronic decay channels then the Bethe-Heitler complication is
irrelevant.

Our calculation \cite{Schaefer2} starts from a pole fit to Compton
scattering data
over a wide energy and angular range \cite{Wada}. These data are
described by the nucleon Born terms (including exchange) plus the
$s$-channel resonance terms where a coherent summation over as
many as 26 nucleon resonances is performed.

In \cite{Schaefer2} we have generalized this semi-empirical
description
to the case of
outgoing massive, time-like photons; in this way we make sure
that the calculations reproduce the data when the photon
approaches the on-shell point. The coupling to the new
longitudinal degree of freedom of the massive photon is
determined by explicitly demanding current conservation and gauge
invariance.

The analysis by Wada et al. involves momentum-dependent vertex
factors
for the photon-nucleon vertex. We interprete this dependence (on
the three-momentum) as a half-off-shell dependence of the
formfactors for a real photon; in this way we have effectively
determined
the half-off-shell behavior of the formfactors from experiment. In
order to check the consequences of an additional VMD-like
dependence on the photon momentum we then multiply the outgoing
vertex factor by a VMD formfactor that peaks at the vector meson
mass. This procedure assumes that the nucleon
off-shell effects decouple from the photon off-shell effects on
the formfactor.

\begin{figure}
\vspace{10cm}
\caption{Dilepton invariant mass spectrum for $\gamma + p$ at 1.2 GeV
for forward (left) and backward (right) emission angles of the
dilepton pair
(from \protect\cite{Schaefer2}).}
\end{figure}

The results of these calculations are shown in Figs. 4. The left part
of Fig. 4
gives the dilepton invariant mass spectrum for forward emission
angles
of the dilepton (small opening angles of the pair with respect to the
direction of the incoming photon momentum). It is evident that the
spectrum is dominated over a wide range by the Bethe-Heitler
contribution.
Nevertheless the spectrum shows a shoulder-like structure in the
vector
meson mass region, which is dominated by the resonance channels. For
large
opening angles, i.e. backward emission (right part of Fig. 4), the
overall
cross section is smaller,
but the resonance structure is more pronounced; again the resonances
contribute somewhat more to it than the nucleon Born graph. Thus, an
experiment of this sort will determine a formfactor that
is averaged over the nucleon and its excited states.

\section{Conclusions}

In this contribution I have discussed a novel approach to the
electromagnetic formfactor of the nucleon in the timelike regime.
I have shown that existing data on proton induced dilepton
production on light nuclei seem to be in gross disagreement with
predictions based on a naive VMD picture. Based on results of a
microscopic model I have argued that
this disagreement may be due to off-shell effects on the
formfactors. I have also indicated that it may be difficult to
disentangle the effects of the strong interaction and the
electromagnetic
interaction vertices in nucleon-nucleon collisions.

An experiment that would give direct access to the time-like
elektromagnetic formfactor without the strong interaction
complications is Compton scattering into the time-like region. I
have presented calculations and have shown that the dilepton
spectra expected from such a reaction are quite sensitive to the
electromagnetic formfactor.

\end{document}